\documentclass[sigconf]{acmart}

\AtBeginDocument{%
  \providecommand\BibTeX{{%
    \normalfont B\kern-0.5em{\scshape i\kern-0.25em b}\kern-0.8em\TeX}}}

\usepackage{listings}
\usepackage{xcolor}

\setcopyright{acmlicensed}
\copyrightyear{2026}
\acmYear{2026}
\acmDOI{XXXXXXX.XXXXXXX}
\acmConference[SIGIR '26]{Proceedings of the 49th International ACM SIGIR Conference on Research and Development in Information Retrieval}{July 12--16, 2026}{Melbourne, VIC, Australia}
\acmISBN{978-1-4503-XXXX-X/18/06}

\renewcommand\shortauthors{}

\graphicspath{{./images/}} 

\begin{document}

\title{GenFacet: End-to-End Generative Faceted Search via Multi-Task Preference Alignment in E-Commerce}

\author{Zhouwei Zhai}
\email{zhaizhouwei1@jd.com}
\authornotemark[1]
\affiliation{%
  \institution{JD.com}
  \city{BeiJing}
  \country{China}}

\author{Min Yang}
\email{yangmin.aurora@jd.com}
\affiliation{%
  \institution{JD.com}
  \city{BeiJing}
  \country{China}}

\author{Jin Li}
\email{lijin.257@jd.com}
\affiliation{%
  \institution{JD.com}
  \city{BeiJing}
  \country{China}}

\begin{abstract}
Faceted search acts as a critical bridge for navigating massive e-commerce catalogs, yet traditional systems rely on static rule-based extraction or statistical ranking, struggling with emerging vocabulary, semantic gaps, and a disconnect between facet selection and underlying retrieval. In this paper, we introduce GenFacet, an industrial-grade, end-to-end generative framework deployed at JD.com. GenFacet reframes faceted search as two coupled generative tasks within a unified Large Language Model: Context-Aware Facet Generation, which dynamically synthesizes trend-responsive navigation options, and Intent-Driven Query Rewriting, which translates user interactions into precise search queries to close the retrieval loop. To bridge the gap between generative capabilities and search utility, we propose a novel multi-task training pipeline combining teacher-student distillation with GRPO. This aligns the model with complex user preferences by directly optimizing for downstream search satisfaction. Validated on China's largest self-operated e-commerce platform via rigorous offline evaluations and online A/B tests, GenFacet demonstrated substantial improvements. Specifically, online results reveal a relative increase of 42.0\% in facet Click-Through Rate (CTR) and 2.0\% in User Conversion Rate (UCVR). These outcomes provide strong evidence of the benefits of generative methods for improving query understanding and user engagement in large-scale information retrieval systems.
  
\end{abstract}




\maketitle

\section{Introduction}
Faceted Search serves as a critical interactive paradigm bridging the gap between users' ambiguous intents and massive item repositories. By providing a ``click-and-filter'' mechanism based on attribute dimensions, it offers a minimalist interactive experience for general users and acts as a pivotal bridge for precise localization~\cite{wei2013survey,yee2003faceted}. By transforming unstructured queries into structured attribute selections, faceted search significantly reduces cognitive load and enhances retrieval efficiency, making it a standard feature on mainstream platforms such as Amazon and JD.com.

Existing technologies primarily fall into two categories: static extraction methods based on rules or external knowledge bases (\texttt{e.g., utilizing WordNet~\cite{stoica2007automating} or Wikipedia~\cite{dakka2008automatic}}), and statistical dynamic ranking methods, such as the DFDRF framework proposed by Pradhan et al. ~\cite{pradhan2023dynamic}. However, in open-domain or large-scale Web environments, traditional Boolean filtering logic is often overly strict and inefficient, failing to cope with query diversity ~\cite{kong2014extending}. Furthermore, the Business Intelligence (BI) aggregation approach introduced by Ben-Yitzhak et al. ~\cite{benyitzhak2008beyond} faces significant computational challenges when handling high-dimensional, sparse data in modern e-commerce.

Current systems face three core limitations in complex modern e-commerce scenarios:
\begin{itemize}
\item Insufficient Timeliness and Adaptability: Methods relying on static lexicons or statistics ~\cite{wei2013survey,yee2003faceted} struggle to capture emerging trend attributes (\texttt{e.g., ``Dopamine Dressing''}) and implicit long-tail attributes. This results in filtering options that lag behind rapid shifts in user cognition.
\item Persistent ``Vocabulary Gap'': A semantic disconnection exists between users `colloquial queries and systems' structured attributes. While Manku et al. ~\cite{manku2021shoptalk} introduced ``intent operators'' in ShopTalk, they still rely on expensive and delayed manual mapping rules, making automated generalization difficult.
\item Disconnection of Personalized Intent in the Retrieval Chain: Traditional three-stage architectures (Retrieval-Facet Generation-Ranking) suffer from severe cascading errors ~\cite{lu2024session}. Personalization work by Koren et al. ~\cite{koren2008personalized} focuses solely on ``what facets to display'' while neglecting subsequent retrieval logic. Users' filtering actions (explicit feedback) fail to propagate back to the underlying retrieval model. Consequently, refined choices are not reflected in the final item retrieval, creating a rift between interaction and retrieval.
\end{itemize}
The core of these challenges lies in: How to achieve end-to-end semantic alignment among unstructured user intents, dynamic personalized behaviors, and structured item knowledge graphs while guaranteeing industrial-grade low latency?

To this end, we propose GenFacet, an LLM-based, end-to-end generative faceted search framework tailored for industrial e-commerce scenarios. Its core idea is to reconstruct faceted search into two tightly coupled generative tasks: Context-Aware Facet Generation and Intent-Driven Query Rewriting. Through multi-task Supervisory Fine-Tuning (SFT)~\cite{ouyang2022training} and further post-training with GRPO (Group Relative Policy Optimization)~\cite{shao2024deepseekmath}, we align these tasks with final search utility, unifying them within the semantic space of a single LLM.

The main contributions of this paper are as follows:
\begin{itemize}
\item We propose the GenFacet framework: The first end-to-end generative framework to unify faceted search into context-aware facet generation and intent-driven query rewriting, effectively resolving semantic gaps and pipeline disconnections.
\item We design an RL-based multi-task alignment method: Innovatively combining teacher-student distillation with Group Relative Policy Optimization (GRPO)~\cite{shao2024deepseekmath}, we directly associate user filtering interactions with the optimization of search result satisfaction, achieving deep alignment from ``displaying filters'' to ``precise retrieval''.
\item We demonstrate significant industrial impact: Extensive offline evaluations on real-world data from the JD.com platform show significant improvements. Online A/B testing reveals a significant 42\% relative increase in Facet Click-Through Rate (CTR) and a 2\% relative increase in User Conversion Rate (UCVR) for facet users, validating the method's effectiveness in real-world e-commerce search scenarios. The system has been fully deployed on JD Search.
\end{itemize}
%% ---------------------------------------------------
%% 2. The GenFacet Framework
%% ---------------------------------------------------

\section{The GenFacet Framework}

\subsection{Problem Formulation \& Overview}

Formally, let $\mathcal{Q}$ denote the space of user queries and $\mathcal{C}$ represent the comprehensive context, including user profile $u_{prof}$, real-time behavior $u_{beh}$, product knowledge graph $\mathcal{G}$, and external web knowledge $\mathcal{W}$. The traditional faceted search problem is often treated as a static classification task. In contrast, we reformulate it as a conditional generation problem. Our goal is to learn a parameterized function $\pi_\theta$ that performs two coupled tasks: 
\begin{enumerate}
    \item \textbf{Facet Generation}: mapping $(q, \mathcal{C})$ to a ranked list of dynamic facets $\mathcal{F} = \{f_1, \dots, f_k\}$
    \item \textbf{Query Rewriting}: mapping a user's selection $f_{sel} \in \mathcal{F}$ back to an optimized retrieval query $q'$ that maximizes the downstream search utility function $U(q')$.
\end{enumerate}

\begin{figure}[h!]
  \centering
  \includegraphics[width=\linewidth]{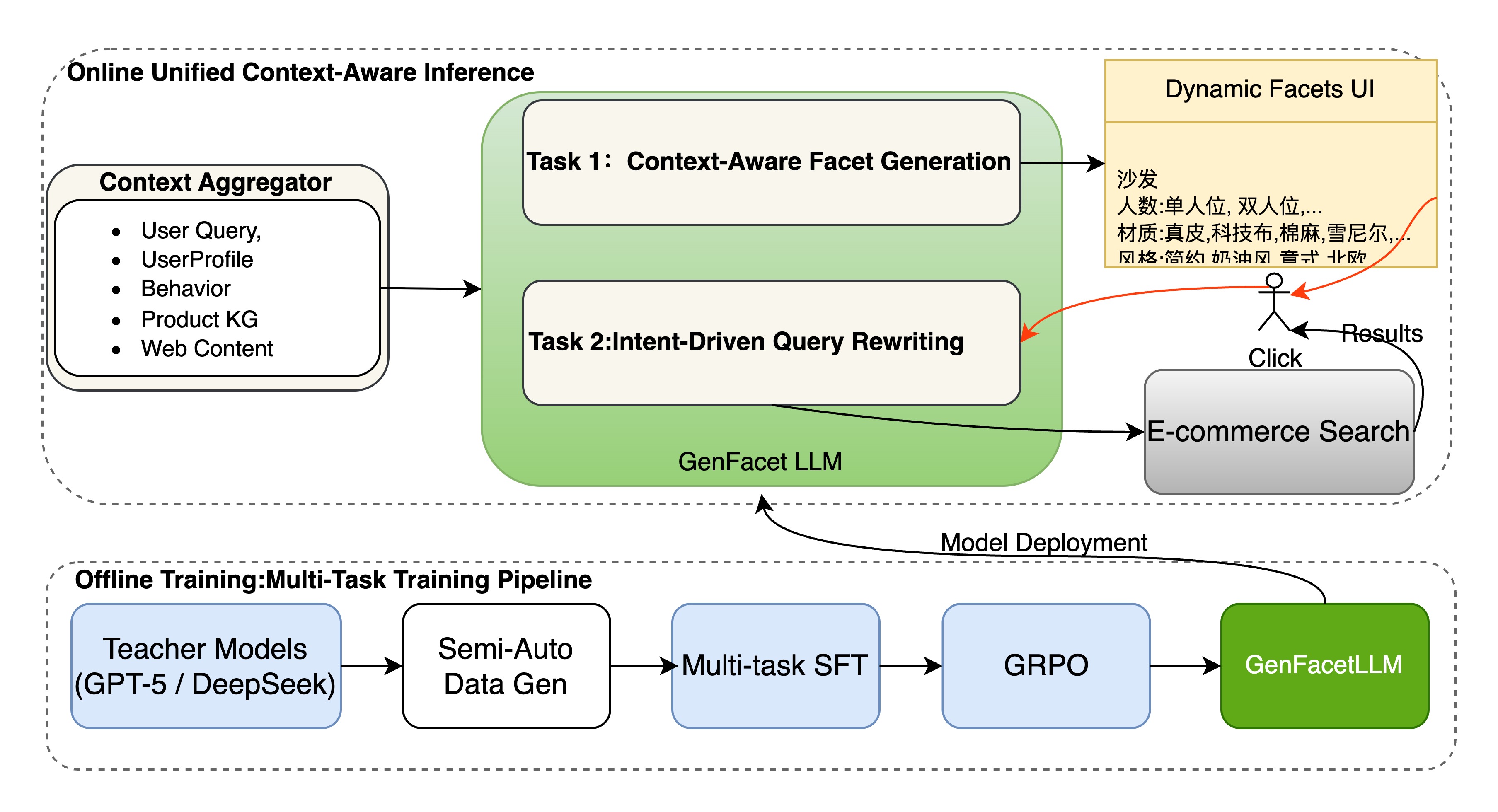}
  \caption{the GenFacet Overview}
  \label{fig:overview}
\end{figure}

Framework Overview. Figure~\ref{fig:overview} illustrates the GenFacet architecture. The system operates in two phases. In the Unified Context-Aware Inference phase, GenFacet aggregates multi-modal context (Query, Profile, Behavior, KG, Web Content) into a structured prompt. This input drives the GenFacetLLM to generate contextually relevant facets (Task 1). Upon user interaction, the specific selection—combined with the interaction history—triggers the model to generate a rewritten query (Task 2), which is then executed by the underlying search engine to retrieve precise results, effectively closing the loop between attribute selection and document retrieval. To ensure robustness, we design a Multi-Task Training Pipeline comprising three stages: Teacher-Student Distillation using reasoning-heavy models (\texttt{e.g., DeepSeek-R1~\cite{guo2025deepseekr1}}), Multi-Task Supervised Fine-Tuning (SFT)~\cite{ouyang2022training}, and preference alignment via Group Relative Policy Optimization (GRPO)~\cite{shao2024deepseekmath} to directly optimize search utility.
\subsection{Unified Context-Aware Inference}
GenFacet unifies facet generation and query rewriting within a single Large Language Model (LLM) inference process, ensuring semantic consistency across the search session.

\textbf{Context-Aware Facet Generation.} To address the ``vocabulary gap'' and emerging trends (\texttt{e.g., ``Dopamine dressing''}), we construct a rich context vector $\mathbf{x}_{ctx}$. This vector aggregates the raw query $q$, user profile interests $u_{prof}$, session-based click/cart behaviors $u_{beh}$, local sub-graphs from the Product KG $\mathcal{G}$ related to $q$, and real-time Web Content $\mathcal{W}$ retrieved via search APIs. These inputs are serialized into a prompt template $T_{gen}$. The probability of generating a sequence of facets $\mathcal{F}$ is given by:
$$P(\mathcal{F} | q, \mathcal{C}) = \prod_{t=1}^{T} P(y_t | y_{<t}, T_{gen}(q, u_{prof}, u_{beh}, \mathcal{G}, \mathcal{W}); \theta)$$
where $y_t$ represents tokens in the generated facet list. This generation allows for dynamic attribute extraction that transcends static indices, capturing latent user needs. 

The prompt is as follows:
\begin{lstlisting}[linewidth=\linewidth]
You are an AI assistant for an e-commerce search system. Based on the user's search context, generate a list of relevant facets (like product attributes) that can help them refine their search.
User Query: {query}
User Profile Interests: {user_profile}
User Session Behaviors (clicks/carts): {user_behavior}
Related Product Knowledge Graph: {kg_subgraph}
Real-time Web Trends: {web_content}
Generate a list of facets in JSON format, each with a name and possible values. Focus on attributes that are not obvious from the query alone and reflect current trends or specific user needs.
Facets:
\end{lstlisting}

\textbf{Intent-Based Filtering Query Rewriting.} Standard systems treat facets as rigid Boolean filters (\texttt{e.g., AND category=``dress''}). However, this often leads to zero-recall or overly narrow results for long-tail queries. GenFacet treats the user's click on a facet value $f_{sel}$ as an explicit relevance feedback signal. We construct a rewriting context $\mathbf{x}_{rw}$ combining the original query, the selected facet, and the interaction history. The model then generates a rewritten query $q'$:
$$q' = \operatorname*{arg\,max}_{q^*} P(q^* | q, f_{sel}, \mathcal{H}_{click}; \theta)$$
Unlike existing approaches~\cite{ma2023query,liu2024query,peng2024large}, this rewritten query $q'$ is semantically aligned with the user's refined intent and is sent to the retrieval engine. By updating the retrieval query rather than merely filtering a static candidate set, GenFacet ensures that the top-ranked results dynamically align with the user's clarified intent, significantly improving recall for complex, high-dimensional demands. 

The prompt is as follows:
\begin{lstlisting}
You are an AI assistant for an e-commerce search system. A user has just clicked on a facet to refine their search. Rewrite the original query to better reflect their new intent for the retrieval engine.
Original Query: {original_query}
Selected Facet: {selected_facet_value} (from facet: {selected_facet_name})
User Click History in this session: {click_history}
Generate ONLY the rewritten query string that captures the user's refined intent. 
Rewritten Query:
\end{lstlisting}

\subsection{Multi-Task Training Pipeline}
\label{sec:subsection}

To adapt a general-purpose LLM to the strict latency and accuracy requirements of industrial e-commerce, we propose a three-stage training paradigm.

\subsubsection{Stage 1: Teacher-Student Distillation}

Since high-quality labels for ``ideal facets'' or ``perfect rewrites'' are unavailable in click logs, we employ strong reasoning models (GPT-5 and DeepSeek-R1) as teachers. We sample 50k instances of real user traffic (queries + context). The teacher models generate candidate facet lists and rewritten queries via chain-of-thought prompting. These outputs are verified by human experts to ensure domain correctness, creating a distilled dataset $\mathcal{D}_{distill}$.

\subsubsection{Stage 2: Multi-Task Supervised Fine-Tuning (SFT)}
We fine-tune the student model (GenFacetLLM) to minimize the joint negative log-likelihood of both tasks. Let $\mathcal{L}_{gen}$ and $\mathcal{L}_{rw}$ be the losses for facet generation and query rewriting, respectively. The total loss is defined as:
$$\mathcal{L}_{SFT} = \mathbb{E}_{\mathcal{D}_{distill}} [-\log P(\mathcal{F}| \mathbf{x}_{ctx}) - \lambda \log P(q'| \mathbf{x}_{rw})]$$
where $\lambda$ balances the task weights. This embeds both capabilities into a shared semantic space.

\subsubsection{Stage 3: Preference Alignment via GRPO}
SFT alone cannot optimize for the non-differentiable metric of ``search utility.'' We employ Group Relative Policy Optimization (GRPO)~\cite{shao2024deepseekmath} to align the model with search satisfaction. Unlike PPO~\cite{schulman2017proximal} which requires a value model, GRPO estimates baselines from group scores, reducing computational overhead.
We define task-specific reward functions: $R_{facet}$ measures the Facet Coverage and predicted CTR of generated attributes; $R_{query}$ executes the rewritten query $q'$ in a pseudo-search environment to measure Recall and Semantic Relevance. The objective is to maximize:
$$\mathcal{J}(\theta) = \mathbb{E}_{q \sim \mathcal{D}, \{o_i\}_{i=1}^G \sim \pi_{\theta_{old}}} \left[ \frac{1}{G} \sum_{i=1}^G \frac{\pi_\theta(o_i|q)}{\pi_{\theta_{old}}(o_i|q)} \hat{A}_i \right] - \beta \mathbb{D}_{KL}(\pi_\theta || \pi_{ref})$$
where $\hat{A}_i$ is the advantage computed from normalized rewards $(R_{facet} + R_{query})$ within the group $G$. This ensures the model favors outputs that not only look plausible but actually drive search conversion.

%% ---------------------------------------------------
%% 3. Industrial Deployment
%% ---------------------------------------------------

\section{Industrial implementation at JD.com}

\subsection{Efficient Model Serving \& Inference}
We address the computational bottleneck by applying knowledge distillation~\cite{hinton2015distilling} to compress the teacher model into a lightweight Qwen3-4B backbone (GenFacetLLM). To accelerate inference, we employ INT8 quantization~\cite{jacob2018quantization} and speculative decoding~\cite{chen2023accelerating}, alongside optimized KV-cache~\cite{liu2024scissorhands} management. Furthermore, to tackle the ``freshness'' issue of emerging attributes, the system integrates the Baidu Search API for real-time external knowledge retrieval, ensuring the model captures long-tail and trending concepts (\texttt{e.g., ``Dopamine Dressing''}) unavailable in static training data.

\subsection{Deployment \& Data Flywheel}
The system is deployed on dual NVIDIA H800 GPU nodes. We introduced a session-aware caching mechanism to handle multi-turn user interactions efficiently. These optimizations reduced average latency to 400ms for facet generation and 180ms for query rewriting, meeting the platform’s strict Service Level Agreements (SLAs).
Crucially, we established a closed-loop data flywheel. Post-deployment, explicit user feedback (clicks on generated facets) and implicit signals (add-to-cart actions) are harvested to form preference pairs. This data drives a continuous iterative training loop, allowing GenFacet to dynamically adapt to shifting user intents and maximize long-term search utility.

%% ---------------------------------------------------
%% 4. Experiments
%% ---------------------------------------------------

\section{Experiments}

\subsection{Experimental Setup}
\subsubsection{Dataset.}
We constructed JD-Facet, a dataset derived from desensitized search logs at JD.com. It contains 5,000 search sessions covering diverse domains. For each session, domain experts annotated the ``ideal facet list'' and relevance scores for the retrieved products after facet selection, serving as the ground truth.
\subsubsection{Baselines.}
We compare GenFacet against three distinct paradigms:
\begin{enumerate}
    \item Rule-based (Production): The current JD.com online system, which relies on a static ``Category-Attribute'' knowledge graph ~\cite{stoica2007automating}. 
    \item DFDRF ~\cite{pradhan2023dynamic}: A dynamic facet ranking framework utilizing Gini coefficients and historical interaction signals. 
    \item Qwen3-4B (Zero-shot): An LLM-based approach using prompting without domain-specific fine-tuning or alignment.
\end{enumerate} 
\subsubsection{Metrics.} 
\begin{enumerate}
    \item Facet Quality: Precision@10 (P@10) and Recall@10 (R@10) measure the accuracy and coverage of the generated facets. 
    \item Retrieval Effectiveness: nDCG@10 evaluates the final product ranking after the complete interaction cycle (Generation $\to$ Interaction $\to$ Rewriting $\to$ Retrieval). 
    \item Online Metrics:Facet CTR (Click-Through Rate) and UCVR (User Conversion Rate).
\end{enumerate}

\subsection{Offline Performance}
Table~\ref{tab:offline_perf} presents the comparative results. GenFacet consistently outperforms all baselines across all metrics.

Comparison with Baselines.

The Rule-based method exhibits reasonable precision ($0.852$) due to curated knowledge bases but suffers significantly in recall ($0.584$) as it fails to capture long-tail or emerging attributes (\texttt{e.g., ``dopamine dressing''}). Conversely, the zero-shot Qwen3-4B demonstrates higher recall ($0.715$) than rules but drops sharply in precision ($0.631$) due to hallucinations invalid for the specific inventory. GenFacet achieves the best of both worlds ($P@10=0.920$, $R@10=0.847$), proving that SFT effectively grounds the LLM in the product space.

Critically, in the Retrieval (nDCG@10) metric, GenFacet achieves a substantial gain ($+15.1\%$ over Rule-based). This confirms that bridging the ``vocabulary gap'' via Intent-Driven Query Rewriting is superior to hard boolean filtering used by traditional methods.

Ablation Study.

We analyze three variants to validate our components:
\begin{enumerate}
    \item w/o GRPO: Removing the reinforcement learning alignment leads to a minor drop in facet metrics but a significant decay in $nDCG@10$ ($0.783 \to 0.741$). This indicates that GRPO is crucial for aligning the generation probability with the utility of the search results (i.e., generating facets that actually lead to good items).
    \item w/o Multi-task SFT: Training tasks independently degrades performance significantly, confirming that the shared semantic space benefits both facet generation and query rewriting.
    \item w/o Query Rewriting: Replacing the rewriter with simple boolean filtering (appending the facet text to the query) causes the largest drop in retrieval performance ($nDCG@10=0.712$). This highlights that the semantic gap between user selection and index terms is the primary bottleneck in traditional systems.
\end{enumerate}

\begin{table}[htbp]
\centering
\caption{Offline Performance Comparison \& Ablation Study.
\textit{Relative improvements are calculated against the Rule-based production baseline.}}
\label{tab:offline_perf}
\footnotesize
\setlength{\tabcolsep}{4pt} % 调整列间距
\begin{tabular}{lccc}
\toprule
\textbf{Method} & \textbf{Facet Gen P@10} & \textbf{Facet Gen R@10} & \textbf{Retrieval nDCG@10} \\
\midrule
Rule-based & 0.852 & 0.584 & 0.680 \\
DFDRF & 0.795 & 0.642 & 0.705 \\
Qwen3-4B (0-shot) & 0.631 & 0.715 & 0.612 \\
GenFacet & 0.920 & 0.847 & 0.783 \\
\textit{Imprv. vs. Production} & \textit{+7.9\%} & \textit{+45.0\%} & \textit{+15.1\%} \\
\midrule
\multicolumn{4}{l}{\textbf{Ablation Variants}} \\
w/o GRPO & 0.895 & 0.812 & 0.741 \\
w/o Multi-task SFT & 0.764 & 0.730 & 0.695 \\
w/o Query Rewriting & 0.918 & 0.845 & 0.712 \\
\bottomrule
\end{tabular}
\end{table}

\subsection{Online A/B Testing}
We deployed GenFacet on the JD App search platform and conducted an A/B test on 10\% of live traffic over two weeks against the production Rule-based system.
\begin{itemize}
    \item \textbf{Engagement:}The Facet \textbf{CTR} increased by a significant \textbf{42.0\%}(relative, p<0.05). This validates that context-aware facets are far more attractive and relevant to users than static category trees.
    \item \textbf{Conversion:}More importantly, the \textbf{UCVR} (User Conversion Rate) for users who interacted with facets increased by \textbf{2.0\%} (relative, p<0.05). Given the massive scale of JD.com, this improvement represents a substantial uplift in Gross Merchandise Value (GMV). This confirms that GenFacet does not just encourage ``clicks'', but effectively guides users to their desired products through the closed-loop intent alignment.
\end{itemize}

%% ---------------------------------------------------
%% 5. Conclusion
%% ---------------------------------------------------

\section{Conclusion}
In this work, we presented GenFacet, the first end-to-end generative framework designed to revolutionize faceted search in large-scale e-commerce environments. By transitioning from rigid extraction to dynamic generation, GenFacet effectively addresses the challenges of adaptability, vocabulary gaps, and the fragmentation between user interaction and retrieval logic. Our proposed multi-task training paradigm, enhanced by GRPO-based preference alignment, successfully unifies context-aware facet generation and intent-driven query rewriting into a single, cohesive semantic space.

The deployment of GenFacet at JD.com provides compelling evidence of its industrial viability. The substantial improvements in Facet CTR (+42.0\%) and UCVR (+2.0\%) confirm that capturing user intent through generative interactions translates directly into superior business outcomes. GenFacet not only enhances the ``searchability'' of complex product catalogs but also establishes a new standard for next-generation, intent-aligned information retrieval systems. 

Future work targets three key advancements: incorporating multimodal inputs for visual facet generation, modeling long-term user sequences for lifetime-aware personalization, and leveraging model quantization for further reducing inference latency. 

\subsection*{Presenter Biography}
\textbf{Zhouwei Zhai} is a Scientist at JD.com, focusing on LLM-powered search systems and AI Agents. At JD.com, he spearheaded the transition towards LLM-augmented e-commerce search and led the end-to-end construction of the platform's next-generation AI Search Assistant.

\bibliographystyle{ACM-Reference-Format}
\bibliography{references}

@article{wei2013survey,
  title={A Survey of Faceted Search},
  author={Wei, Bifan and Liu, Jun and Zhu, Qiaoming},
  journal={Journal of Web Engineering},
  volume={12},
  number={1 \& 2},
  pages={041--064},
  year={2013}
}

@inproceedings{yee2003faceted,
  title={Faceted metadata for image search and browsing},
  author={Yee, Ka-Ping and Swearingen, Kirsten and Li, Kevin and Hearst, Marti},
  booktitle={Proceedings of the SIGCHI Conference on Human Factors in Computing Systems (CHI '03)},
  pages={401--408},
  year={2003},
  publisher={ACM}
}

@inproceedings{pradhan2023dynamic,
  title={Dynamic Filter Discovery and Ranking Framework for Search and Browse Experiences in E-Commerce},
  author={Pradhan, Ligaj and Yu, Le and Malvar, Sara},
  booktitle = {Proceedings of the 2023 ACM SIGIR Workshop on eCommerce},
  year={2023}
}

@inproceedings{stoica2007automating,
  title={Automating Creation of Hierarchical Faceted Metadata Structures},
  author={Stoica, Emilia and Hearst, Marti A and Richardson, Megan},
  booktitle={Proceedings of the Human Language Technology Conference of the NAACL (NAACL-HLT '07)},
  pages={244--251},
  year={2007}
}

@inproceedings{dakka2008automatic,
  author={Dakka, Wisam and Ipeirotis, Panagiotis G.},
  title={Automatic Extraction of Useful Facet Hierarchies from Text Databases}, 
  year={2008},
  booktitle = {Proceedings of the 2008 IEEE International Conference on Data Engineering (ICDE)},
  pages = {466--475},
  doi={10.1109/ICDE.2008.4497455}
}

@inproceedings{koren2008personalized,
    author = {Koren, Jonathan and Zhang, Yi and Liu, Xue},
    title = {Personalized interactive faceted search},
    year = {2008},
    isbn = {9781605580852},
    publisher = {Association for Computing Machinery},
    address = {New York, NY, USA},
    url = {https://doi.org/10.1145/1367497.1367562},
    doi = {10.1145/1367497.1367562},
    pages = {477–486},
    numpages = {10},
    booktitle = {Proceedings of the 17th International World Wide Web Conference},
    series = {WWW '08}
}

@inproceedings{lu2024session,
  title={Session-aware product filter ranking in e-commerce search},
  author={Lu, Hanqing and Tang, Xianfeng and Luo, Chen and Cui, Limeng and Dai, Zhenwei and Goutam, Rahul and Zhang, Haiyang and Cheng, Monica Xiao},
  booktitle={The Second Tiny Papers Track at ICLR 2024},
  year={2024},
  url={https://openreview.net/forum?id=r4LMF2IJ6R}
}

@inproceedings{benyitzhak2008beyond,
    author = {Ben-Yitzhak, Ori and Golbandi, Nadav and Har'El, Nadav and Lempel, Ronny and Neumann, Andreas and Ofek-Koifman, Shila and Sheinwald, Dafna and Shekita, Eugene and Sznajder, Benjamin and Yogev, Sivan},
    title = {Beyond basic faceted search},
    year = {2008},
    isbn = {9781595939272},
    publisher = {Association for Computing Machinery},
    address = {New York, NY, USA},
    url = {https://doi.org/10.1145/1341531.1341539},
    doi = {10.1145/1341531.1341539},
    booktitle = {Proceedings of the 2008 International Conference on Web Search and Data Mining},
    pages = {33–44},
    numpages = {12},
    location = {Palo Alto, California, USA},
    series = {WSDM '08}
}

@article{manku2021shoptalk,
  title={ShopTalk: A System for Conversational Faceted Search},
  author={Manku, Gurmeet and Lee-Thorp, James and Kakarla, Ashwin},
  journal = {arXiv},
  note = {arXiv:2109.00702},
  year={2021},
  doi={10.48550/arXiv.2109.00702}
}

@inproceedings{kong2014extending,
  title={Extending Faceted Search to the General Web},
  author={Kong, Weize and Allan, James},
  booktitle={Proceedings of the 23rd ACM International Conference on Information and Knowledge Management (CIKM '14)},
  pages={839--848},
  year={2014},
  publisher={ACM}
}

@article{guo2025deepseekr1,
  title = {DeepSeek-R1: Incentivizing Reasoning Capability in {LLMs} via Reinforcement Learning},
  author = {Guo, Daya and Yang, Dejian and Zhang, Haowei and Song, Junxiao and 
                  Zhang, Ruiyu and Xu, Runxin and Zhu, Qihao and Ma, Shirong and 
                  Wang, Peiyi and Bi, Xiao and others},
  journal = {arXiv},
  year = {2025},
  note = {arXiv:2501.12948},
  doi={10.48550/arXiv.2501.12948}
}

@article{ouyang2022training,
  title={Training language models to follow instructions with human feedback},
  author={Ouyang, Long and Wu, Jeffrey and Jiang, Xu and Almeida, Diogo and Wainwright, Carroll L and Mishkin, Pamela and others},
  journal = {arXiv},
  note = {arXiv:2203.02155},
  year={2022},
  doi={10.48550/arXiv.2203.02155}
}

@article{shao2024deepseekmath,
  title={DeepSeekMath: Pushing the Limits of Mathematical Reasoning in Open Language Models},
  author={Shao, Zhihong and Wang, Peiyi and Zhu, Qihao and Xu, Runxin and Song, Junxiao and Bi, Xiao and others},
  journal = {arXiv},
  note = {arXiv:2402.03300},
  year={2024},
  doi={10.48550/arXiv.2402.03300}
}

@inproceedings{ma2023query,
  title={Query rewriting in retrieval-augmented large language models},
  author={Ma, Xuequn and Gong, Yeyun and He, Pengshan and others},
  booktitle={Proceedings of the 2023 Conference on Empirical Methods in Natural Language Processing},
  pages={5303--5315},
  year={2023}
}

@article{liu2024query,
  title={GenRewrite: Query Rewriting via Large Language Models},
  author={Liu, Jerry and Mozafari, Barzan},
  journal = {arXiv},
  note = {arXiv:2403.09060},
  year={2024},
  doi={10.48550/arXiv.2403.09060}
}

@inproceedings{peng2024large,
  title={Large language model based long-tail query rewriting in taobao search},
  author={Peng, Wei and Li, Guanyu and Jiang, Yan and others},
  booktitle={Companion Proceedings of the ACM Web Conference 2024},
  pages={20--28},
  year={2024}
}

@article{schulman2017proximal,
  title={Proximal Policy Optimization Algorithms},
  author={Schulman, John and Wolski, Filip and Dhariwal, Prafulla and Radford, Alec and Klimov, Oleg},
  journal = {arXiv},
  note = {arXiv:1707.06347},
  year={2017},
  doi={10.48550/arXiv.1707.06347}
}

@article{hinton2015distilling,
  title={Distilling the knowledge in a neural network},
  author={Hinton, Geoffrey and Vinyals, Oriol and Dean, Jeff},
  journal = {arXiv},
  note = {arXiv:1503.02531},
  year={2015},
  doi={10.48550/arXiv.1503.02531}
}

@inproceedings{jacob2018quantization,
  title={Quantization and training of neural networks for efficient integer-arithmetic-only inference},
  author={Jacob, Benoit and Kligys, Skirmantas and Chen, Bo and Zhu, Menglong and Tang, Mingxing and Howard, Andrew and Adam, Hartwig and Kalenichenko, Dmitry},
  booktitle={Proceedings of the IEEE Conference on Computer Vision and Pattern Recognition (CVPR'18)},
  pages={2704--2713},
  year={2018},
  organization={IEEE},
  doi={10.1109/CVPR.2018.00286}
}

@article{chen2023accelerating,
  title={Accelerating large language model decoding with speculative sampling},
  author={Chen, Charlie and Borgeaud, Sebastian and Irving, Geoffrey and Lespiau, Jean-Baptiste and Sifre, Laurent and Jumper, John},
  journal = {arXiv},
  note = {arXiv:2302.01318},
  year={2023},
  doi={10.48550/arXiv.2302.01318}
}

@article{liu2024scissorhands,
  title={Scissorhands: Exploiting the persistence of importance hypothesis for LLM KV cache compression at test time},
  author={Liu, Zichang and Desai, Aditya and Liao, Feng and Sivashankar, Varma and Lu, Yiyang and Cheng, Xiang and Tian, Yuanyuan and Zhao, Ashley and Zhuang, Liang and Shieh, S and Chen, P},
  journal = {arXiv},
  note = {arXiv:2305.17118},
  year={2023},
  doi={10.48550/arXiv.2305.17118}
}

\end{document}